\documentclass[10pt,twocolumn]{IEEEtran}

\usepackage{amsmath}
\usepackage{amsthm}
\usepackage{amssymb}
\usepackage{graphicx}
\usepackage{subfig}
\usepackage{color}
\newsavebox{\tempbox}

\newcommand{\bad}{\boldsymbol{!}}

\newcommand{\abs}[1]{\vert #1 \vert }

\newcommand{\Pt}[1]{P_{#1}}

\newcommand{\defn}{\triangleq}

\newcommand{\typicaleps}[2]{T^{n}_{\left[ #1 \right]_{#2} }}

\newcommand{\Ipi}[1]{I\left( #1  \right)}
\newcommand{\cb}[1]{\mathbf{c}^n_{#1}}
\newcommand{\CBS}[1]{\mathbf{C}^n_{#1}}

\newcommand{\msf}[1]{\mathsf{#1}}

\newcommand{\mcf}[1]{\mathcal{#1}}\usepackage{color}
\newcommand{\idc}[1]{\mathfrak{1}\left({#1}\right) }

\newtheorem{theorem}{Theorem}
\newtheorem{define}{Definition}

\begin{document}

\title{A coding approach to guarantee information integrity against a
  Byzantine relay}

\author{Eric Graves and Tan F. Wong \\
  Department of Electrical \& Computer Engineering \\
  University of Florida, FL 32611 \\
\texttt{\{ericsgra,twong\}@ufl.edu}}

\maketitle

\begin{abstract}
 This paper
  presents a random coding scheme with which two nodes can exchange
  information with guaranteed integrity over a two-way Byzantine
  relay. This coding scheme is employed to obtain an inner bound on
  the capacity region with guaranteed information integrity.  No
  pre-shared secret or secret transmission is needed for the proposed
  scheme. Hence the inner bound obtained is generally larger than
  those achieved based on secret transmission schemes.  This approach
  advocates the separation of supporting information integrity and
  secrecy.
\end{abstract}

\section{Introduction}
In order for two parties to communicate through an intermediary node,
it is required that the intermediary (hereto referred to as a relay)
must faithfully forward the information. In a communication network,
such cooperative behavior is not guaranteed as relays may have reasons
for forwarding false information in order to fool the intended
participants.  Such attacks, oft referred to as \emph{Byzantine
  attacks}, have major ramifications on protocols that
operate within the network.  For instance, Maurer~\cite{Maurer97}
addressed the need for authentication in order to support secret key
agreement in a simple two-node system with an active eavesdropper. In
much the same way, integrity of information must be guaranteed for
larger systems to ensure secrecy. The genesis of this problem has
roots in cryptography \cite{bloch2011} where codes
like~\cite{GilbertBSTJ74,AMD} have been studied as a means of
determining Byzantine attacks, while more recent work has focused on
the integrity of a network using linear network coding \cite{KosutALL09, WangTIT10}, as well as the
study of using coding to determine manipulation in basic channels which are supported by a relay \cite{HeISIT09,Mao}.

Historically this problem of guaranteeing information integrity is
treated by the use of a non-observable key to add redundancy to
information that allows attack detection.
In \cite{GilbertBSTJ74}, the problem, originally motivated by a
dishonest gambling pit boss, was studied with use of planar codes and
random codes. While in \cite{AMD}, Cramer et al. show that any random
linear transformation into a space of greater dimension is with high
probability invertible. Thus if one were to modify the transformation
or the symbols, it is with high probability that the modified sequence would not
be one corresponding to any valid input sequence. The
resulting codes are known as algebraic manipulation detection (AMD)
codes.

In order to extend these ideas to the physical layer, additional
redundancy is required to cope with the possibility of channel errors.
Mao and Wu \cite{Mao} posed the problem of trying to determine which
relay in a multiple relay two-hop network was manipulating the data. A
cross-layer method is set forth in which a cryptographic key is
inserted into the signal, by which the intended destination determines
from the physical layer error rate if manipulation has occurred.
In slight contrast, the more recent work of He and Yener
\cite{HeISIT09}, mainly focused on the problem in the two-way two-hop
channel studied in this paper, does not require use of a shared secret
key. Instead, an LDPC code is employed to support secret transmission
which in turns allows the use of an AMD code to detect attacks by the relay key.
The major drawback of this solution is not separating the need for
secrecy and integrity. As a result, it does not provide a deeper
understanding of what is actually needed to support the two different
requirements, making extensions to beyond the addition channel
considered in \cite{HeISIT09} difficult.

In contrast we propose a different strategy by separating the concepts
of secrecy and integrity to ensure that communication can be verified
without use of any key or any secret transmission. Using random coding
techniques, we obtain an inner bound on the capacity region with
guaranteed information integrity in the general scenario of two
nodes which must communicate through a Byzantine relay node.
Section~\ref{se:model} describes the channel model in detail. The
inner bound on the capacity region with guaranteed information
integrity is provided in Section~\ref{se:main}. An outline of the
achievability proof that leads to the inner bound is given in
Section~\ref{se:achievability}. The notation employed in the paper is
summarized in Table~\ref{tb:notation}.

\section{Two-way Amplify-and-forward Relay Model} \label{se:model}

\begin{figure}
\centering 
\subfloat[Multiple-access
channel]{\includegraphics[width=0.48\textwidth]{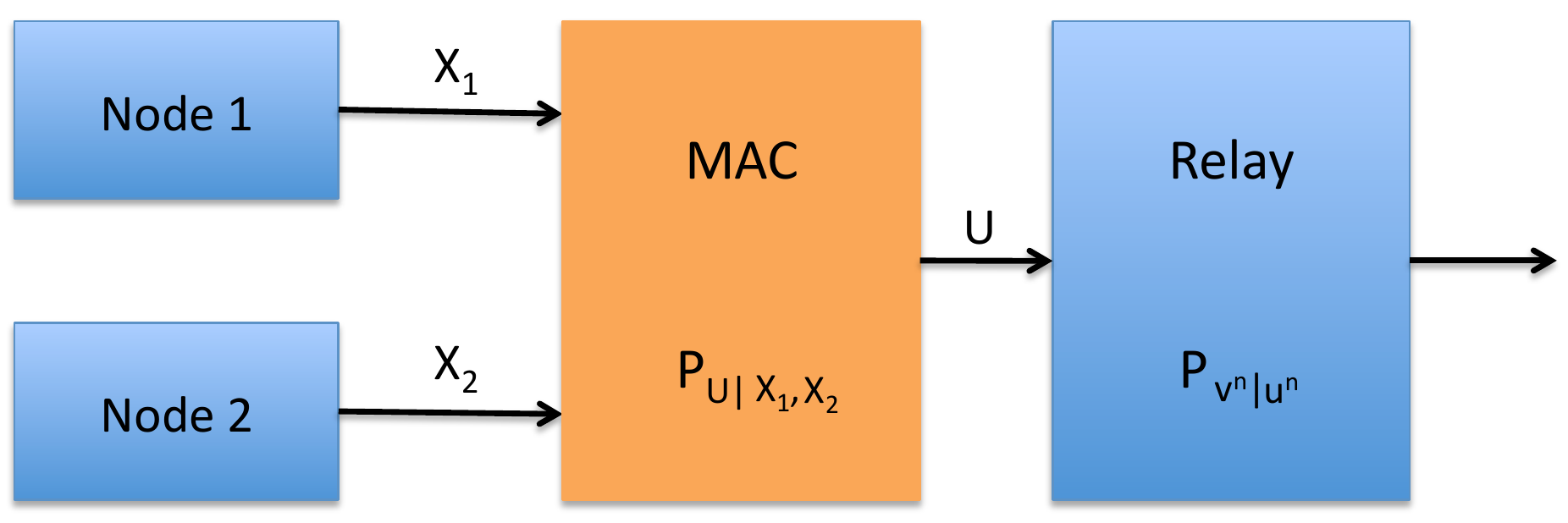}
  \label{fig:model1}} 
\hfil
\subfloat[Broadcast channel]{
    \includegraphics[width=0.48\textwidth]{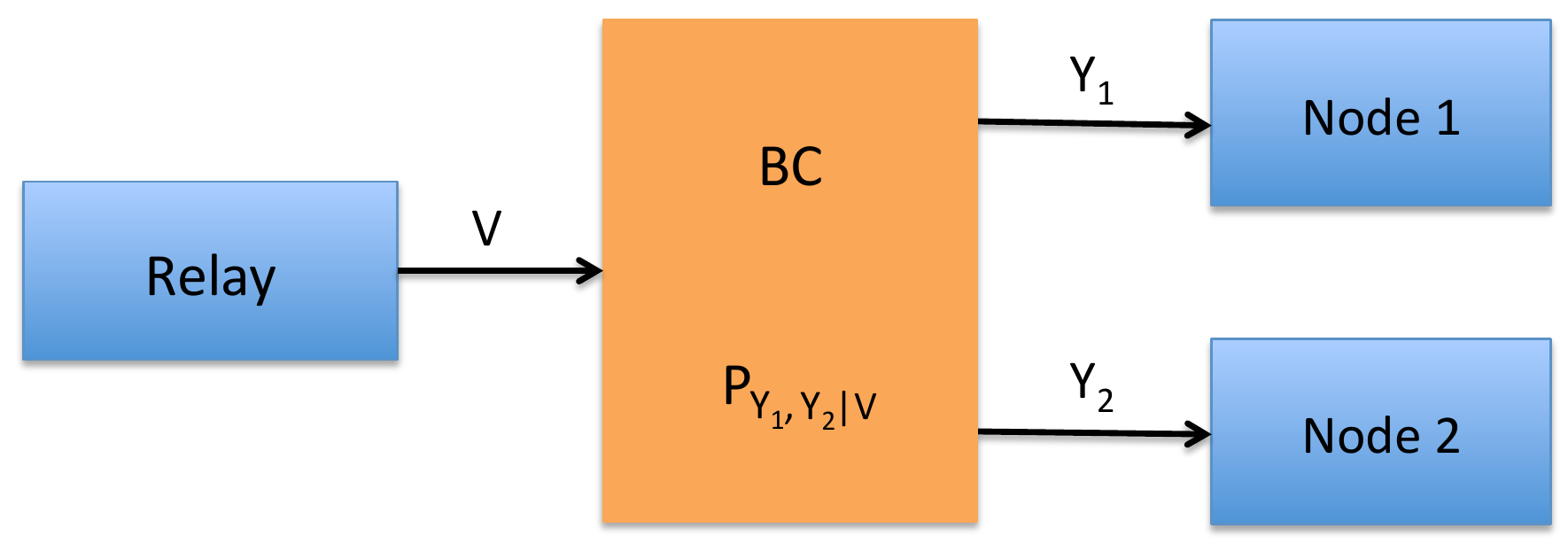}
    \label{fig:model2}}
  \caption{Two-way, half-duplex, amplify-and-forward relay model.}
\label{fig:model}
\end{figure}

Consider the two-way, half-duplex relay channel model shown in
Fig.~\ref{fig:model},
in which two nodes (1 and 2) simultaneously send symbols to a relay
node through a discrete, memoryless multiple-access channel (MAC).
The half-duplex relay node is then supposed to broadcast its received
symbols back to the two nodes in the amplify-and-forward (AF)
manner. For simplicity, we assume that the broadcast channel (BC) from
the relay back to the nodes is perfect. That is, both nodes perfectly
observe the symbols sent out by the relay.  There is some possibility
that the relay may modify its received symbols in an attempt to
conduct a manipulation attack. The design goal is for each node to at
least detect any malicious act of the relay in the event that it can
not decode the information sent by the other node; in other word, to
guarantee the integrity of the information forwarded by the relay.

\begin{table}
\centering
\caption{Notation}
\label{tb:notation}
\begin{tabular}{|c c|}
\hline
$X$ & random variable \\
$\msf{x} \in \mcf{X}$ & the element $\msf{x}$ from the alphabet of $X$ \\
$x^n$ & $n$ instances of random variable $X$ over $\mcf{X}$ \\
$N(\msf{x} \vert x^n)$ & number of times $\msf{x}$ occurs in $x^n$\\
$ \Pt{x^n}(\msf{x})$ & $\frac{1}{n}N(\msf{x} \vert x^n)$ \\
$\Pt{x^n \vert y^n}(\msf{x}\vert \msf{y})$ & $\frac{\Pt{x^n,y^n}(\msf{x},\msf{y})}{\Pt{y^n}(\msf{y})}$ \\
$[X]^n_\delta $ & $\big\{ \Pt{x^n} : \left| \Pt{x^n}(\msf{x}) - P(\msf{x})
\right| \leq \delta  \big\}$ \\
$\typicaleps{X}{\delta} $ & $ \{x^n : \Pt{x^n} \in [X]^n_\delta\} $\\
$\Pt{X|Y}(\msf{x}|\msf{y})$ & cond. pmf of $X$ given $Y$; also
treated as a matrix \\
$\idc{\cdot}$ & indicator function \\
\hline 
\end{tabular}
\end{table}
More specifically, let $\mathcal{X}_1$, $\mathcal{X}_2$, and
$\mathcal{U}$ denote the discrete alphabets of node 1's input, node
2's input, and the output of the MAC.  Over time instants
$1,2,\ldots,n$, suppose that nodes 1 and 2 transmit the symbol
sequences $x_1^n$ and $x_2^n$, respectively, through the memoryless
MAC. The output sequence $U^n$ is conditionally distributed according
to
\begin{equation}\label{eq:mac}
p(u^n| x_1^n,x_2^n) = \prod_{i=1}^n \Pt{U|X_1,X_2}(u_i|x_{1,i},x_{2,i})
\end{equation}
where the conditional pmf
$\Pt{U|X_1,X_2}(\msf{u}|\msf{x}_1,\msf{x}_2)$ specifies the MAC.  The
relay node, during time instants $1,2,\ldots,n$, observes the output
symbol sequence $U^n$ of the MAC, processes (or manipulates) it, and
then broadcasts the processed symbol sequence to nodes 1 and 2 at time
instants $n+1,n+2,\ldots,2n$ via the perfect BC.  Let $\mcf{V}$ be the
alphabet of the relay's processed symbols. Because the relay is
supposed to work in the AF manner, there must be a one-to-one
correspondence between the elements of $\mcf{U}$ and $\mcf{V}$. Thus,
without loss of generality, we may assume that $\mcf{V}=\mcf{U}$.  Let
$V^n$ denote the relay's output sequence.  The assumption of perfect
BC from the relay to the nodes implies that $\mcf{Y}_1= \mcf{Y}_2 =
\mcf{U}$, $\Pt{Y_1|V} = \Pt{Y_2|V} = I$, and
\begin{equation}
p(y_1^n,y^n_2|v^n) = \idc{y_1^n = y_2^n = v^n}.
\label{eq:bc}
\end{equation}
For convenience hereafter, we simply make $V^n$ the symbol sequence
observed by both nodes.

Fix $n$. Let $R_1$ and $R_2$ be two positive rates. Consider the
encoder-decoder quadruple $(\CBS{1},\CBS{2},g^n_1,g^n_2)$:
\begin{align*}
  \CBS{1} &: \{1,2,\ldots,2^{nR_1}\} \rightarrow \mathcal{X}^n_1 \\
  \CBS{2} &: \{1,2,\ldots,2^{nR_2}\} \rightarrow \mathcal{X}^n_2 \\
  g^n_1 &: \mathcal{U}^n \times \{1,2,\ldots,2^{nR_1}\} \rightarrow
  \{1,2,\ldots,2^{nR_2}\} \cup
  \{\bad\}  \\
  g^n_2 &: \mathcal{U}^n \times \{1,2,\ldots,2^{nR_2}\} \rightarrow
  \{1,2,\ldots,2^{nR_1}\} \cup \{\bad\}
\end{align*}
where $\CBS{1}$ and $g^n_1$ are the encoder and decoder used by
node~1, and $\CBS{2}$ and $g^n_2$ are the encoder and decoder used by
node~2. Note that we allow the encoders $\CBS{1}$ and $\CBS{2}$ to be
random. The symbol $\bad$ in the decoder output alphabets denotes the
decision that the received sequence is deemed untrustworthy, i.e., the
relay has possibly been malicious.  Let $W_1$ and $W_2$ be independent
messages of nodes~1 and 2 that are uniformly distributed over
$\{1,2,\ldots,2^{nR_1}\}$ and $\{1,2,\ldots,2^{nR_2}\}$,
respectively. Then $X_1^n = \CBS{1}(W_1)$ and $X_2^n = \CBS{2}(W_2)$
are the codewords sent by nodes~1 and 2 to the relay through the MAC.

The potential manipulation by the relay is specified by the
conditional distribution of $V^n$ given the other random quantities
mentioned above.  We impose the Markovity restriction on the
conditional distribution that
\begin{equation}\label{eq:relay}
p(v^n|u^n, x_1^n, x_2^n, w_1, w_2, \cb{1}, \cb{2}) = p(v^n|u^n,
\cb{1}, \cb{2})
\end{equation}
which means that the relay may potentially manipulate the transmission
based only on the output symbols of the MAC that it observes as well
as its knowledge about the codebooks used by the nodes.  If
$p(v^n|u^n, \cb{1}, \cb{2}) = \idc{v^n=u^n}$, then we regard the relay
as \emph{non-malicious}. Otherwise the relay is malicious. For later
presentation clarity, we will employ $H_1$ and $H_0$ to denote the
conditions that the relay is and is not malicious, respectively.

With the scheduling and coding scheme described above, we say that the
rate pair $(R_1,R_2)$ is \emph{achievable with guaranteed information
  integrity} if there exists a sequence of encoder-decoder quadruples
$\{(\CBS{1},\CBS{2},g^n_1,g^n_2)\}$ such that:
\begin{align*}
  \text{Under } H_0 &: \\
  & \hspace{-40pt} \Pr \left\{ g^n_1(V^n,W_1) \neq W_2 \cup
    g^n_2(V^n,W_2) \neq W_1 \right\} \rightarrow 0 \\
  \text{Under } H_1 &: \\
  & \hspace{-40pt}\Pr \left\{ g^n_1(V^n,W_1) \notin \{W_2,\bad\}
    \cup g^n_2(V^n,W_2) \notin \{W_1,\bad\} \right\}
  \rightarrow 0
\end{align*}
as $n \rightarrow \infty$. Note that the requirement under $H_1$
forces the decoders to either detect the substitution attack by the
relay or correct the symbols modified.  The \emph{capacity region with
  guaranteed information integrity} in this case can then be defined
as the closure of the set of achievable rate pairs with guaranteed
information integrity.  Note that we have not counted the use of the
perfect BC from the relay back to the nodes in the rate definition
above. If that is to be counted, the factor of $0.5$ should be added
to all rates because the fixed transmission schedule descrived above.

Note that from \eqref{eq:mac}, \eqref{eq:bc}, and \eqref{eq:relay},
the joint distribution of $(V^n,U^n,X^n_{1}, X^n_2,W_{1},
W_2,\CBS{1},\CBS{2})$ is given by
\begin{align}
& \hspace{-10pt} p(v^n, u^n, x_1^n, x_2^n, w_1, w_2, \cb{1}, \cb{2})
\notag \\
  &= p(v^n\vert u^n, \cb{1}, \cb{2}) \, p(u^n \vert x_1^n, x_2^n) \,
  \idc{\cb{1}(w_{1})=x_{1}^n} \notag \\
  &\hspace{10pt} \cdot \idc{\cb{2}(w_{2})=x_{2}^n} \, p(w_{1})
  \, p(w_{2}) \, p(\cb{1},\cb{2}).
\label{eq:maindistr}
\end{align}



\section{Inner bound on capacity region}  \label{se:main}

Under the operational definition of information transfer with
guaranteed integrity given in Section~\ref{se:model}, it turns out
that the matrix-algebraic structure of \emph{manipulability} given in
\cite{GravesIT12} is critical to our inner bound on the capacity
region. For easy reference, we repeat the definition of manipulability
for node~1's \emph{observation channel} specified by the stochastic
matrix pair $(P_{U|X_1},P_{Y_1|V})$ in the notation of this paper:
\begin{define}\label{def:manipulable}
  The observation channel $(P_{U|X_1},P_{Y_1|V})$ is manipulable if
  there exists a $\abs{\mcf{U}}\!\times\!\abs{\mcf{U}}$ non-zero
  matrix $\Upsilon$, whose $j$th column, for each
  $j=1,2,\ldots,\abs{\mcf{U}}$, is balanced and $(0,0)$-polarized at
  $j$, with the property that all columns of $\Upsilon P_{U|X_1}$ are
  in the right null space of $P_{Y_1|V}$.  Otherwise,
  $(P_{U|X_1},P_{Y_1|V})$ is said to be non-manipulable.
\end{define}
Manipulability of node~2's observation channel $(P_{U|X_2},P_{Y_2|V})$
is the same.  

\begin{theorem}\label{thm:main}
  An inner bound on the capacity region with guaranteed information
  integrity is the closure of the convex hull of all $(R_1,R_2)$
  satisfying
\begin{align*}
  R_1 & < I(X_1; U|X_2) \\
  R_2 & < I(X_2; U|X_1)
\end{align*}
for some 
\[
P_{U,X_1,X_2}(\msf{u},\msf{x}_1,\msf{x}_2) =
P_{U|X_1,X_2}(\msf{u}|\msf{x}_1,\msf{x}_2) P_{X_1}(\msf{x}_1)
P_{X_2}(\msf{x}_2)
\]
having the property that $(P_{U|X_1},I)$ and $(P_{U|X_2},I)$ are both
non-manipulable.
\end{theorem}
Note that the difference between the above region and the standard
capacity region (without guaranteed integrity) is that the former does
not contain the rate pairs generated by input distributions that do
not give non-manipulable observation channels while the latter does.

\begin{figure}
\centering
\includegraphics[width=0.3\textwidth]{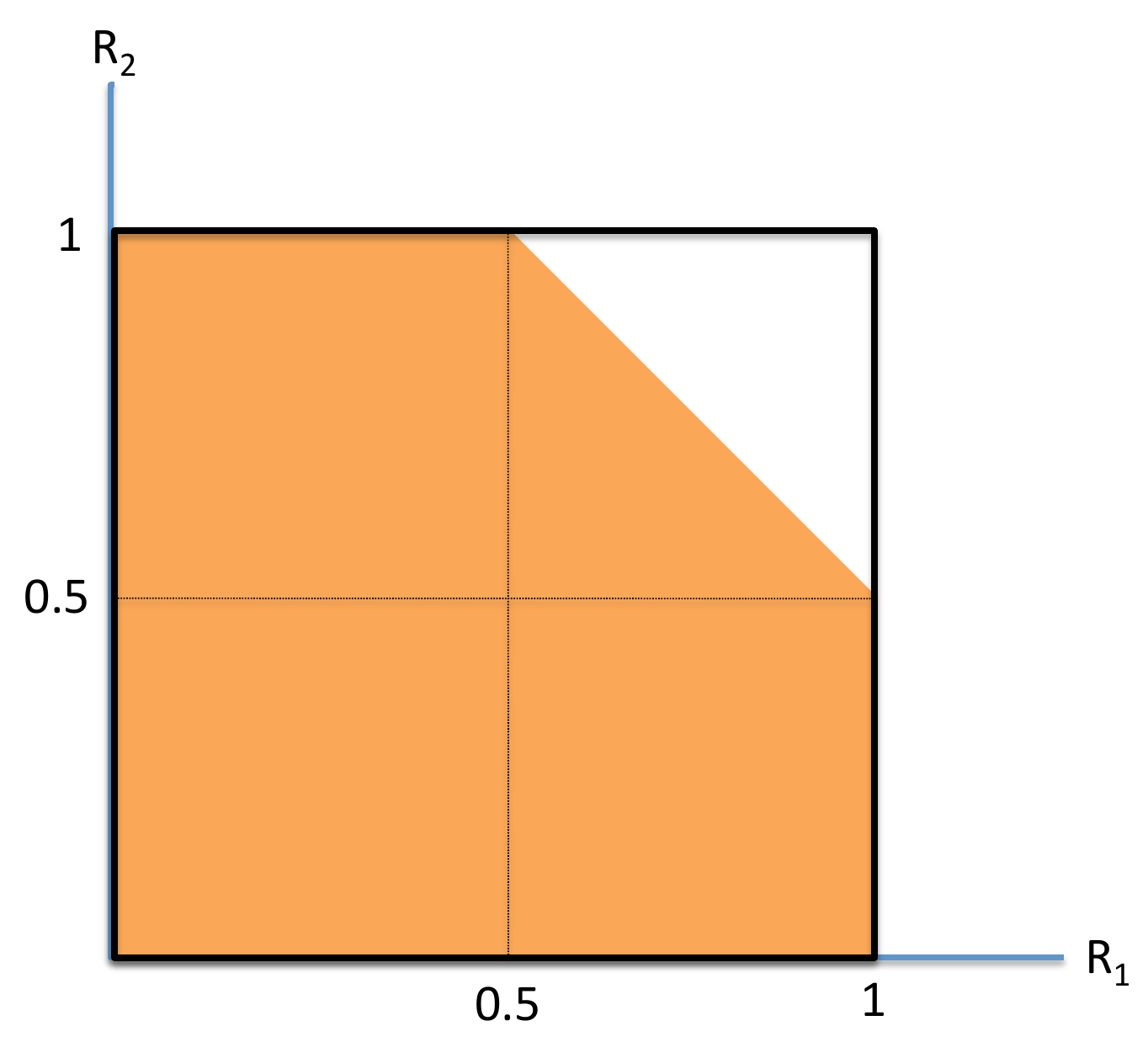}
\caption{Capacity region with guaranteed information integrity of the
  two-way AF relay channel with a binary erasure MAC and a perfect
  BC.}
\label{fig:2waycap}
\end{figure}

Let us apply Theorem~\ref{thm:main} to the simple example two-way AF
relay channel made up of a binary erasure MAC \cite{Cover2006} and a
perfect BC. That is, $X_1$ and $X_2$ have the same binary alphabet
$\{0,1\}$. The MAC is described by $U=X_1+X_2$. Hence the alphabet of
$U$ and $V$ are both $\{0, 1, 2\}$. The BC is defined by $Y_1=V$ and
$Y_2=V$, and $P_{Y_1|V} = P_{Y_2|V} = I$.  Let $P_{X_1}(1)=p$ and
$P_{X_1}(1)=q$, where $0 \leq p,q \leq 1$. Then
\[
P_{U|X_1} = \left( \begin{matrix}
1-q &0 \\
q & 1-q \\
0& q
\end{matrix} \right)
\text{~~and~~} 
P_{U|X_2} = \left( \begin{matrix}
1-p &0 \\
p & 1-p \\
0& p
\end{matrix} \right).
\]
It is not hard to check \cite[Thm.~2]{GravesIT12} that both
$(P_{U|X_1},I)$ and $(P_{U|X_2},I)$ are non-manipulable for all
choices of $p$ and $q$.  Thus the inner bound in
Theorem~\ref{thm:main} implies that the capacity regions with and
without guaranteed information integrity are identical as shown by the
thick-lined square region in Fig.~\ref{fig:2waycap}, and can be
achieved by choosing both $X_1$ and $X_2$ to be equally likely binary
random variables.
With this choice of input distributions, the shaded area shown in
Fig.~\ref{fig:2waycap} is the capacity region of the binary erasure
MAC channel, treating the relay as the MAC receiver. From the
achievability argument in Section~\ref{se:achievability}, to guarantee
integrity while operating in the shaded region, we need randomization
in the encoders to move the operating point up to the unshaded
triangular portion of the capacity region. Physically, this means that
we need to confuse the relay from decoding both nodes' messages. Note
that for this two-way relay channel, uncoded transmission can achieve
capacity without guaranteed information integrity, while coded
transmission is needed in order to achieve capacity with guaranteed
information integrity. We point out that this two-way relay channel is
also considered in \cite{HeISIT09}, our result shows that attack
detection can be indeed achieved without secrecy.

\section{Outline of Achievability} \label{se:achievability}

We employ the standard random coding argument to show that the
probabilities of both error events under $H_0$ and $H_1$ employed in
the definition of achievable rate pairs in Section~\ref{se:model},
averaged over all random codebooks, converge to zero as $n$ increases
under the conditions stated in Theorem~\ref{thm:main}.  Then the
existence of a codebook pair (and the corresponding decoding
functions) having the same property is guaranteed.

\subsection{Code Construction}

Fix $\Pt{X_1}(\msf{x}_1)$ and $\Pt{X_2}(\msf{x}_2)$ that satisfy the
condition of non-manipulable $(\Pt{U|X_1},I)$ and $(\Pt{U|X_2},I)$.
It can then be shown that $I(X_1;U) < I(X_1;U|X_2)$ and $I(X_2;U) <
I(X_2;U|X_1)$.  If $I(X_1;U) < R_1 < I(X_1;U|X_2)$, $I(X_2;U) < R_2 <
I(X_2;U|X_1)$, and $R_1+R_2 > I(X_1,X_2;U)$, independently and
uniformly pick $2^{nR_1}$ codewords $\CBS{1}(1), \CBS{1}(2), \ldots,
\CBS{1}(2^{nR_1})$ from the typical set $\typicaleps{X_1}{\delta_n}$,
where $\{\delta_n\}$ satisfies the delta convention set forth in
\cite{CK}. Similarly, pick $2^{nR_2}$ codewords $\CBS{2}(1),
\CBS{2}(2), \ldots, \CBS{2}(2^{nR_2})$ from the typical set
$\typicaleps{X_2}{\delta_n}$.  Instantiations of $\CBS{1}$ and
$\CBS{2}$ define the deterministic encoding functions for nodes~1 and
2, respectively. If $(R_1,R_2)$ does not satisfy the above conditions,
choose another pair $(R'_1,R'_2)$, with $R'_1 \geq R_1$ and $R'_2 \geq
R_2$, that does.  Randomly pick $2^{nR'_1}$ and $2^{nR'_2}$ codewords
from $\typicaleps{X_1}{\delta_n}$ and $\typicaleps{X_2}{\delta_n}$ as
above to form codebooks for nodes~1 and 2, respectively. Pick
independent (of all other random quantities) random numbers $W'_1$
uniformly from $\{1,2,\ldots, 2^{n(R'_1-R_1)}\}$ and $W'_2$ uniformly
from $\{1,2,\ldots, 2^{n(R'_2-R_2)}\}$. Then employ the random
encoding function which maps $W_1$ to $\CBS{1}\left((W_1-1)
  2^{n(R'_1-R_1)} + W'_1\right)$, and the random encoding function
which maps $W_2$ to $\CBS{2}\left((W_2-1) 2^{n(R'_2-R_2)} +
  W'_2\right)$. Clearly if we can decode to $(W_1-1) 2^{n(R'_1-R_1)} +
W'_1$, we can also obtain $W_1$, and if we can decode to $(W_2-1)
2^{n(R'_2-R_2)} + W'_2$, we can also obtain $W_2$. Therefore, we may
simply assume below $(R_1,R_2)$ satisfy $I(X_1;U) < R_1 <
I(X_1;U|X_2)$, $I(X_2;U) < R_2 < I(X_2;U|X_1)$, and $R_1+R_2 >
I(X_1,X_2;U)$, and employ the deterministic encoders without loss of
any generality.

For a fixed codebook pair $(\cb{1},\cb{2})$, nodes~1 employs (and
symmetrically for node 2) the following typicality decoder:
\begin{align*}
g^n_1(v^n,w_1) &= 
\begin{cases} {w}_{2} & \text{if } 
  (v^n,\cb{1}(w_1),\cb{2}({w}_{2})) \in
  \typicaleps{U, X_1, {X}_{2}}{2\mu_n}  \\
  & \text{and there is no } 
  \hat{w}_2 \neq w_2 \text{ such that} \\
  & (v^n,\cb{1}(w_1),\cb{2}({\hat w}_{2})) \in
  \typicaleps{U, X_1, {X}_{2}}{2\nu_n}, \\
  \bad & \text{otherwise},
\end{cases} 
\end{align*}
where $\mu_n$ is a function of $\delta_n$ and the alphabets $\mcf{U}$,
$\mcf{X}_1$, and $\mcf{X}_2$, and $\nu_n \geq \mu_n$ will be specified
later. Both $\mu_n$ and $\nu_n$ satisfy the delta convention. Below we
will also make use of $\mu'_n$, $\mu''_n$, $\tilde \mu_n$, and $\hat
\mu_n$, which are all constant multiples of $\mu_n$.


\subsection{Error analysis under $H_0$}

Under this case, we have $V^n = U^n$.  
Because of symmetry and the union bound, it suffices to consider
$\Pr\left\{ g^n_1(V^n,W_1) \neq {W}_{2} \right\}$.  To that end,
define
\begin{align*}
\mathcal{U} (x_1^n,x_2^n) & \defn
\left\{ u^n : (u^n,x_1^n,x_2^n) \in \typicaleps{U, {X}_{1},X_2}{2\mu_n}
\right\}, \\
\mathcal{V}_{\beta}(x_1^n;w_2,\cb{2}) & \defn  \\
& \hspace*{-30pt} \left\{ u^n :
  \bigcup_{{\hat w}_{2} \neq {w}_{2}} \hspace{-5pt} 
  (u^n,x_1^n,\cb{2}({\hat w}_{2})) \cap
  \typicaleps{U, X_{1}, {X}_{2}}{\beta} \neq \emptyset \right\},
\end{align*}
respectively.  Then
\begin{align}
\Pr\left\{ g^n_1(U^n,W_1) \neq {W}_{2}\right\} 
  & \leq
  \Pr\{ U^n \notin \mathcal{U}(\CBS{1}(W_1),\CBS{2}(W_2)) \} \notag \\
  & \hspace{-98pt} + \Pr \big\{U^n \in 
    \mathcal{U}(\CBS{1}(W_1), \CBS{2}(W_2)) \notag \\
  & \hspace{-28pt} \cap
    \mathcal{V}_{2\nu_n}(\CBS{1}(W_1);W_2,\CBS{2}) \big\} .
\label{eq:C0}
\end{align}
It remains to show that both probabilities on the right hand side of
\eqref{eq:C0} converge to zero as $n \rightarrow \infty$.

By the combination of \cite[Problem~2.9]{CK}, \eqref{eq:mac}, and
\cite[Lemma~2.12]{CK}, it is shown that
\begin{align}
  & \hspace*{-10pt}
  \Pr\{ U^n \notin \mathcal{U}(\CBS{1}(W_1),\CBS{2}(W_2)) \}  \notag \\
  &\leq 
  (n+1)^{2\abs{\mcf{X}_1}\abs{\mcf{X}_2}} \, 2^{-n\delta_n} +
  2\abs{\mcf{U}}\abs{\mcf{X}_1}\abs{\mcf{X}_2} \, e^{-2n\delta_n^2}.
\label{eq:C0_A}
\end{align}
Additionally, using the standard argument
(cf. \cite[Ch. 7]{Cover2006}), based on the fact that the codewords in
the codebooks are chosen independently, it is easy to establish
\begin{align*}
& \hspace{-10pt} 
\Pr\{U^n \in \mathcal{U}(\CBS{1}(W_1),\CBS{2}(W_2))
  \cap \mathcal{V}_{2\nu_n}(\CBS{1}(W_1);W_2,\CBS{2}) \} \notag \\
& \leq 2^{-n[I(X_{2}; U \vert X_{1}) -  R_2 - \epsilon_n ]},
\end{align*}
for some $\epsilon_n \rightarrow 0$.

\subsection{Error analysis under $H_1$}
Again by symmetry and the union bound, it suffices to show that
$\Pr\left\{ g^n_1(V^n,W_1) \notin \{{W}_{2},\bad\}\right\}$ vanishes
as $n$ increases.  Allowing $\lambda_n$ to be set later, define
\begin{align*}
E_{1} &=\{(u^n, v^n): \min \Ipi{\tilde X_1;\tilde V \vert \tilde U}  >
\lambda_n  \}\\
E_{2} &=\{(u^n,v^n): \min \Ipi{\tilde X_1; \tilde V \vert \tilde U}
\leq \lambda_n \}
\end{align*}
where for each pair of $(u^n,v^n)$, the minimization of mutual
information above is over all triples $(\tilde X_1,\tilde U, \tilde
V)$ of random variables, which respectively take values over
$\mathcal{X}_1$, $\mathcal{U}$, and $\mathcal{U}$, and have
distributions satisfying the following constraints:
\begin{align}
&\Pt{\tilde U \tilde V}(\msf{u}, \msf{v}) = \Pt{u^n v^n}(\msf{u},\msf{v})
\notag \\ 
&\left| \Pt{\tilde X_1 \vert \tilde U} (\msf{x}_1 \vert \msf{u})-
  \Pt{X_1 \vert U}(\msf{x}_1\vert \msf{u}) \right| \leq \tilde\mu_n 
\notag \\ 
&\left| \Pt{\tilde X_1 \vert \tilde V} (\msf{x}_1 \vert  \msf{v}) -
  \Pt{X_1 \vert U}(\msf{x}_1\vert \msf{v}) \right| \leq \tilde\mu_n.
\label{eq:sh3}
\end{align}

Note that for $(u^n,v^n) \in E_2$, $\min I(\tilde X ; \tilde V \vert
\tilde U) \leq \lambda_n$. To simplify notation, let $(\tilde X,
\tilde U, \tilde V)$ be the choice that achieves the minimum. Then by
the Pinsker inequality \cite[Lemma~11.6.1]{Cover2006}, there exists a
constant $k$ such that
\[
\left| \Pt{\tilde V|\tilde U}(\msf{v} \vert \msf{u}) - \Pt{\tilde V |
    \tilde U, \tilde X_1} (\msf{v} \vert \msf{u}, \msf{x}_1) \right|
\leq k\sqrt{\lambda_n}.
\]
This inequality itself also implies that there exists another constant 
$k'$ such that 
\[
\left|\sum_{u} \Pt{\tilde V| \tilde U}(\msf{v} \vert u) 
  \Pt{U|X_1}(u \vert \msf{x}_1) - \Pt{U|X_1} (\msf{v} \vert \msf{x}_1)
\right| < k'\sqrt{\lambda_n}.
\]
This situation though is analyzed in \cite{GravesIT12}, and can be
shown to imply that there exists another scalar $k''$ giving $\left|
  \Pt{\tilde V|\tilde U}(\msf{v} \vert \msf{u}) - \idc{\msf{v} =
    \msf{u}} \right| \leq k'' \sqrt{\varepsilon}$, provided that
$(\Pt{U|X_1},I)$ is non-manipulable.
Consequently, it can be shown that $(v^n,\cb{1}(w_1), \cb{2}(w_2)) \in
\typicaleps{U,X_1,X_2}{2\tilde k \sqrt{\lambda_n}}$ if $u^n \in
\mathcal{U}(\cb{1}(w_1), \cb{2}(w_2))$, for some constant $\tilde k$.
By setting, $\nu_n = \tilde k \sqrt{\lambda_n}$, we can conclude that
$\left\{ U^n \in \mathcal{U}(\CBS{1}(W_1),\CBS{2}(W_2)), (U^n,V^n) \in
  E_2 \right\}$ is not an error event under $H_1$.
Therefore we can bound $\Pr\left\{ g^n_1(V^n,W_1) \notin
  \{{W}_{2},\bad\}\right\}$ as below:
\begin{align}
& \hspace{-8pt}
 \Pr\left\{ g^n_1(V^n,W_1) \notin \{{W}_{2},\bad\}\right\} \notag\\
& \leq
\Pr\{ U^n \notin \mathcal{U}\left(\CBS{1}(W_1),\CBS{2}(W_2)\right) \} \notag \\
&\hspace{7pt} + \Pr\{ U^n \in \mathcal{U}\left(\CBS{1}(W_1),\CBS{2}(W_2) \right), \notag \\
& \hspace{25pt}
  V^n \in \mathcal{V}_{2\mu_n}(\CBS{1}(W_1);W_2,\CBS{2}), \,
  (U^n,V^n) \in E_1\} 
\label{eq:C1}
\end{align}

The probability $\Pr\{ U^n \notin
\mathcal{U}\left(\CBS{1}(W_1),\CBS{2}(W_2)\right) \}$ has been shown
to vanish above (cf. \eqref{eq:C0_A}). It thus remains to show that
the second probability on the right hand side of \eqref{eq:C1}
decreases to zero as $n \rightarrow \infty$.


To that end, define the following sets for convenience:
\begin{align*}
& \hspace*{-10pt} \mathcal{Q}_1(u^n,v^n;\cb{1}) \\
  & \defn
\left\{ w_1 : \cb{1}(w_1) \in \typicaleps{X_1|U}{\tilde\mu_n}(u^n)
    \bigcap \typicaleps{X_1|U}{\tilde\mu_n}(v^n)
  \right\}, \\
& \hspace*{-10pt} \mathcal{Q}_2(u^n,v^n;\cb{1},\cb{2}) \\
& \defn
\left\{ w_2 : \cb{2}(w_2) \in \hspace{-20pt} \bigcup_{w_1 \in
      \mathcal{Q}_1(u^n,v^n;\cb{1})}  \hspace{-20pt}
    \typicaleps{X_2|U,X_1}{\mu''_n} (u^n, \cb{1}(w_1)) \right\}, \\
& \hspace*{-10pt} \mathcal{Q}_3(w_2; u^n;\cb{1},\cb{2}) \\
& \defn
\left\{ w_1 : \cb{1}(w_1) \in
    \typicaleps{X_1|U,X_2}{\mu'_n} (u^n, \cb{2}(w_2)) \right\}.
\end{align*}
By \cite[Problem~2.10]{CK}, there
exists $\zeta_n$ satisfying the delta convention that
\[
\left| \typicaleps{X_1|U}{\tilde\mu_n}(u^n) \cap
  \typicaleps{X_1|U}{\tilde\mu_n}(v^n) \right| \leq 2^{n\left[\max
    H(\tilde X_1 \vert \tilde U,\tilde V) + \zeta_n \right]},
\]
where $\tilde X_1,\tilde U,\tilde V$ are restricted to distributions
that satisfy~\eqref{eq:sh3}. Through judicious use of this and
\cite[p.~409, Lemma 17.9]{CK}, the following bounds can be shown:
\begin{align*}
 & \hspace{-10pt} 
 \Pr\left\{ \left| \mathcal{Q}_{1}(u^n,v^n; \CBS{1}) \right| > 
  \gamma 2^{n \left[ \left| R_1 - \min I(\tilde X_1; \tilde
        U , \tilde V) \right|^+ + \zeta_n +\epsilon_n \right] } \right\} 
 \notag \\
 & \leq 
e^{-\sigma(\gamma )2^{n \left[\left| R_1 - \min I(\tilde X_1;
        \tilde V| \tilde U)\right|^+ + \zeta_n+\epsilon_n\right]}}, \\
& \hspace{-10pt}
\Pr\Bigg\{ \left| \mathcal{Q}_{2}(u^n,v^n; \CBS{1},\CBS{2}) \right| > 
\notag \\
& \hspace{10pt} 
\gamma
  2^{n\left[\left|R_2 - I(X_2;U|X_1) + \left|R_1 - \min I(\tilde X_1;
        \tilde U, \tilde V)\right|^+ \right|^+ + \zeta_n+3\epsilon_n
\right]}, \notag \\
& \hspace{5pt}
\left| \mathcal{Q}_{1}(u^n,v^n; \CBS{1}) \right| \leq  
  \gamma 2^{n \left[ \left| R_1 - \min I(\tilde X_1; \tilde U , \tilde
        V) \right|^+ + \zeta_n +\epsilon_n \right] }
\Bigg\} \notag \\ 
& \hspace{5pt} 
\leq
e^{-\sigma(\gamma) 2^{n\left[\left|R_2 - I(X_2;U|X_1) + \left|R_1 -
      \min I(\tilde X_1; \tilde U, \tilde V)\right|^+
      \right|^+ + \zeta_n+3\epsilon_n \right]} }, \\
& \hspace{-10pt}
\Pr\left\{ \max_{w_2} \left| \mathcal{Q}_3(w_2; u^n;\CBS{1},\CBS{2}) \right|
  > \gamma 2^{2n\epsilon_n} \right\} \notag \\ 
& \hspace{30pt} \leq
  e^{-\sigma(\gamma) 2^{2n\epsilon_n} + n R_2\ln 2},
\end{align*}
for any $\gamma>1$ and $\sigma(\gamma) \defn \gamma \ln \gamma -
\gamma + 1$. 
Hence all the three probabilities above can be bounded by the double
exponential term $e^{-\sigma(\gamma) 2^{n\epsilon'_n}}$ for some
$\epsilon'_n$ with the property that $n\epsilon'_n \rightarrow
\infty$.  Using this and \eqref{eq:maindistr}, we can bound
\begin{align*}
& \hspace{-5pt}
\Pr \bigg\{ U^n \in \mathcal{U}\left(\CBS{1}(W_1),\CBS{2}(W_2) \right), 
\notag \\
&\hspace{20pt}   
V^n \in \mathcal{V}_{2\mu_n}(\CBS{1}(W_1);W_2,\CBS{2}), (U^n,V^n) \in E_1 \bigg \} 
\notag \\ 
&\leq 
2^{-n[H(U \vert X_1 X_2) +R_1+R_2-\epsilon_n]} 
\sum_{u^n,v^n} \idc{(u^n,v^n) \in E_1}  \notag \\
& \hspace{5pt} \cdot
\idc{u^n \in \typicaleps{U}{\hat\mu_n}}
\idc{v^n \in \typicaleps{U}{\hat\mu_n}}
\sum_{\cb{1},\cb{2}} p(v^n \vert u^n,\cb{1},\cb{2}) \notag \\
& \hspace{5pt} \cdot 
\left| \mathcal{Q}_2(u^n,v^n;\cb{1},\cb{2}) \right|
\cdot \max_{w_2} \left| \mathcal{Q}_3(w_2;u^n;\cb{1},\cb{2}) \right|
p(\cb{1},\cb{2})
\end{align*}
\begin{align*}
& \leq  2^{-n[H(U \vert X_1 X_2) +R_1+R_2-\epsilon_n]}
\bigg\{ 2^{n\left[2H(U)+2\epsilon_n\right]} \cdot
  3e^{-\sigma(\gamma) 2^{n\epsilon'_n} } \notag \\
& \hspace{10pt}  + \hspace{-5pt} \sum_{u^n,v^n,\cb{1},\cb{2}} \hspace{-5pt}  
\idc{u^n \in \typicaleps{U}{\hat\mu_n}}
\idc{(u^n,v^n) \in E_1}
\notag \\
& \hspace{10pt} \cdot 
\gamma^2
2^{n\left[\left|R_2 - I(X_2;U|X_1) + \left|R_1 - \min I(\tilde X_1;
        \tilde U, \tilde V)\right|^+ \right|^+ + \zeta_n+5\epsilon_n \right]}
\notag \\
& \hspace{10pt} \cdot 
p(v^n \vert u^n,\cb{1},\cb{2}) \, p(\cb{1},\cb{2}) 
\bigg\}
\notag \\
& \leq 
3e^{-\sigma(\gamma) 2^{n\epsilon'_n} + n \left[ H(U)+I(X_1,X_2;U)
    -R_1-R_2+3\epsilon_n \right] \ln 2 } \notag \\
& \hspace{10pt} + \gamma^2
2^{-n[ R_1+R_2-I(X_1,X_2; U) - \zeta_n - 7\epsilon_n]}
+ \gamma^2
2^{-n \left[ \lambda_n - \zeta_n - 8\epsilon_n \right]}.
\end{align*}
Hence by setting $\lambda_n = 2\zeta_n + 8 \epsilon_n$, the upper bound
 vanishes.



\section{Conclusion}

We have presented an inner bound on the capacity region of which two
nodes can exchange information with guaranteed integrity through a
two-way Byzantine relay. The inner bound is specified by the
non-manipulability property of the channel. The coding scheme that
achieves the bound requires neither any pre-shared secret nor secret
transmission. As a result, the inner bound is generally larger than
any achievable region that is obtained based on secret transmission.
We believe that the inner bound is in fact the capacity region with
guaranteed information integrity. We will prove this converse
statement in an upcoming paper. Finally we point out that the coding
scheme described can be easily modified to support the additional
requirement of secret transmission.
%

\bibliographystyle{ieeetr} 
\end{document}